\documentclass[aps,prl,twocolumn]{revtex4-1}

\usepackage{natbib,graphicx,dcolumn,color,lipsum}

\usepackage{subfigure,amsmath,amssymb,verbatim,moreverb,bm}

\def\be{\begin{equation}}

\def\ee{\end{equation}}

\def\ber{\begin{eqnarray}}

\def\eer{\end{eqnarray}}

\def\sigmav{\mbox{\boldmath $\sigma$}}

\def\br{{\bf r}}

\def\bp{{\bf p}}

\def\bk{{\bf k}}

\begin{document}

\title{Spin Hall effects due to phonon skew scattering}

\author{Cosimo Gorini}

\affiliation{Institut f\"ur Theoretische Physik, Universit\"at Regensburg, 93040 Regensburg, Germany}

\author{Ulrich Eckern}

\affiliation{Institut f\"ur Physik, Universit\"at Augsburg, 86135 Augsburg, Germany}

\author{Roberto Raimondi}

\affiliation{Dipartimento di Matematica e Fisica, Roma Tre University, Via della Vasca Navale 84, 00146 Rome, Italy}

\begin{abstract}
A diversity of spin Hall effects in metallic systems is known to rely on Mott skew scattering.
In this work its high-temperature counterpart, phonon skew scattering, which is expected
to be of foremost experimental relevance, is investigated. In particular, 
the phonon skew scattering spin Hall conductivity is found to be practically $T$-independent 
for temperatures above the Debye temperature $T_D$.  
As a consequence, in Rashba-like systems a high-$T$ linear behavior of the spin Hall angle demonstrates the dominance
of extrinsic spin-orbit scattering only if the intrinsic spin splitting is smaller than
the temperature.
\end{abstract}

\maketitle

The spin Hall effect (SHE) \cite{dyakonov1971,murakami2003,sinova2004,kato2004,wunderlich2005} 
is the generation of a transverse spin current by an applied electric field, 
the spin current polarization being perpendicular to both current and field directions.  
Indeed, a family or related effects exists \cite{Handbook}.
The SHE and its inverse are routinely employed in spin injection or extraction
experiments in a variety of systems \cite{valenzuela2006,vila2007,mosendz2010,niimi2011,obstbaum2014,weiler2014,isasa2014},
and their potential for spintronics applications is becoming ever more evident 
\cite{Liu12,*LiuLee12}.
A crucial issue is the determination of the dominant spin-orbit mechanism responsible for such
effects. In particular, whether this is of {\it intrinsic} origin, 
i.e., connected with the band and/or device structure or geometry, or {\it extrinsic}, 
i.e., due to impurities.  Spin-orbit phenomena are typically complex in their own right, 
mixing charge and spin (magnetic) degrees of freedom in a plethora of ways, 
and standard experimental setups add to such a complexity \cite{jungwirth2012}.
We will see that one of the main phenomenological arguments employed to
discern the dominant spin-orbit mechanism misses the central aspect
of dynamical spin-orbit interaction. The latter describes {\it inter alia}
the direct interaction between the electrons' spin and phonons, and, though it will be the leading
process at experimental temperatures $T \approx 300$ K, 
it has been mostly neglected until now \cite{toelle2014}.

At $T=0$ in  metallic systems there are three main {\it extrinsic} spin-orbit mechanisms:
(i) side-jump \cite{berger1970}, (ii) skew scattering \cite{smit1955}, and (iii) Elliott-Yafet spin
relaxation \cite{elliott1954,*yafet63}.  When a charge current is driven
through a sample, (i) and (ii) give rise to a transverse spin current via the side-jump 
and skew scattering spin Hall conductivities, denoted $\sigma_{\rm sj}^{\rm sH}$ 
and $\sigma_{\rm ss}^{\rm sH}$, respectively.  Elliott-Yafet spin relaxation is typically weak, 
but is needed to ensure the proper analytical behavior of the full
spin Hall conductivity $\sigma^{\rm sH}$ when also {\it intrinsic} spin-orbit interaction 
is present---as is the case in thin films or two-dimensional (2D) electron or 
hole gases \cite{raimondi2009}.  

The above mechanisms have been extensively studied at $T=0$, 
where they arise from electron scattering at static impurities.  
In this case one has (explicitly in 2D) \cite{Engel05,*Tse06}:
\be
\label{sH0}
\sigma_{\rm sj,0}^{\rm sH} = \frac{en}{\hbar}\left(\frac{\lambda}{2}\right)^2 \, ,
\;
\sigma_{\rm ss,0}^{\rm sH} = \left(\frac{\lambda k_F}{4}\right)^2\frac{en}{m} \, 2\pi N_0 v_0 \tau_0
\ee
with $n$ the electron density, $\lambda$ the effective Compton wavelength 
of extrinsic spin-orbit coupling, $v_0$ the scattering amplitude, $k_F$ the Fermi wave vector, 
$N_0=m/2\pi\hbar^2$ the density of states,  $\tau_0$ the elastic scattering time, and $e>0$ the unit charge. 
Equation~\eqref{sH0} shows that the side-jump conductivity is independent of 
the scattering mechanism (at least in simple parabolic bands), whereas the skew scattering one 
is proportional to $\tau_0$, i.e., to the Drude conductivity $\sigma_D = e^2n\tau_0/m = -en\mu$
($\mu = -e\tau_0/m$ is the mobility).  From these $T=0$ results, the $T\neq0$ spin Hall conductivity 
behavior is extrapolated arguing that the skew scattering conductivity should behave as
$\sigma_{\rm ss}^{\rm sH} \propto \mu$, with the proportionality constant depending 
on microscopic details (impurity concentration, $k_F$, etc.), but {\it not} on the temperature.   
Hence the argument goes as follows \cite{hankiewicz2006,vila2007,vignale2010,niimi2011,isasa2014}: 
(i) in high mobility samples skew scattering should dominate, and 
(ii) the spin Hall signal should scale as the mobility with respect to its $T$-dependence.  
On the contrary, the same signal should be $T$-independent in samples 
where the side-jump mechanism is the leading one. 

However, we will see that this simple and appealing phenomenological extrapolation 
from $T=0$ to $T\neq0$ misses a critical feature of high-$T$ skew scattering---which, 
following Ref.~[\onlinecite{isasa2014}], we call ``phonon skew scattering''.  
Namely that for temperatures $T\gtrsim T_D$, with $T_D$ the Debye temperature, 
$\sigma_{\rm ss}^{\rm sH}$ does {\it not} scale as the mobility and rather becomes $T$-independent.
Since typical spin Hall experiments are performed at room temperature in ``soft'' metals such as
Au ($T_D=165$ K), Pt ($T_D=240$ K) or Ta ($T_D=240$ K) 
\cite{vila2007,kimura2007, *seki2008, *liu2011, *hahn2013, obstbaum2014}, this makes distinguishing between side-jump
and skew scattering contributions an even more complicated issue.

The Hamiltonian is (compare Ref.~[\onlinecite{toelle2014}]) 
\be
H = H_0 + \delta V^{\rm ph}(\br,t)
-\frac{\lambda^2}{4\hbar}\sigmav\times\nabla \delta V^{\rm ph}(\br,t)\cdot\bp + H_1^{\rm ph} \, ,
\label{model1}
\ee
where $\sigmav$ is the vector of Pauli matrices. Here
$H_0$ contains the static electronic part of the Hamiltonian, in first-quantized notation
given by
\be
\label{model0}
H_0^{\rm el} = \frac{p^2}{2m} - \frac{\alpha}{\hbar}\sigmav\times\hat{\bf z}\cdot\bp 
+ V_{\rm imp}(\br) - \frac{\lambda^2}{4\hbar}\sigmav\times\nabla V_{\rm imp}(\br)\cdot\bp \, ,
\ee
as well as the standard harmonic phonon contribution: $H_0 = H_0^{\rm el} + H_0^{\rm ph}$.
The second term on the r.h.s.\ of Eq.~\eqref{model0} is a Bychkov-Rashba-like 
intrinsic spin-orbit term \cite{bychkov1984}, which appears at the interface between 
transition metals and insulators/vacuum where inversion symmetry 
is broken \cite{shikin2008,*varykhalov2008,*rybkin2010}.  The potential from static impurities is
denoted $ V_{\rm imp}(\br)$, and $\delta V^{\rm ph}(\br,t)$ stands---classically speaking---for 
the time-dependent potential due to lattice vibrations at $T\neq0$.

The actual calculations employ well-known quantum field theoretical techniques, see below.  
Here we only mention that it is convenient to introduce the phonon field operator \cite{agdbook}
\be
\label{phonon_field}
\hat\varphi (\br)={i}\sum_{\bk}\sqrt{\frac{v_sk}{2V}}
\left( 
\hat b_{\bk}e^{{i}\bk\cdot \br }-{\rm c.c.}
\right) \, ,
\ee
where $\hat b_{\bk}$ and $\hat b_{\bk}^{\dagger}$ are
annihilation and creation operators for longitudinal Debye phonons of momentum $\hbar\bk$,
$v_s$ is the sound velocity, and $V$ the volume (or the area in 2D).
Note that $\hat\varphi (\br)$ corresponds to $v_s\sqrt{\rho}$ times the divergence 
of the ionic displacement, where $\rho$ is the ionic mass density.  As usual, 
the electron-phonon coupling constant will be denoted by $g$ \cite{agdbook}.
Finally, the anharmonic term (3-phonon processes) reads
\be
\label{phonon_anharmonic}
H_1^{\rm ph}= \frac{\Lambda}{3!}\int\,{\rm d}\br\, \hat\varphi^3 (\br) \, .
\ee
In its most general form, there appears a tensor arising from the third derivatives 
of the crystal potential with respect to small displacements \cite{zimanbook2}.
For our purposes, however, it is sufficient to characterize the anharmonicity by
the single parameter $\Lambda$, which is related to the Gr\"{u}neisen parameter $\gamma$ by 
$\Lambda = -\gamma / \rho^{1/2} v_s$; typically $\gamma\approx 2\dots 3$ \cite{zimanbook2}.

\begin{figure}
\begin{center}
\includegraphics[width=3in]{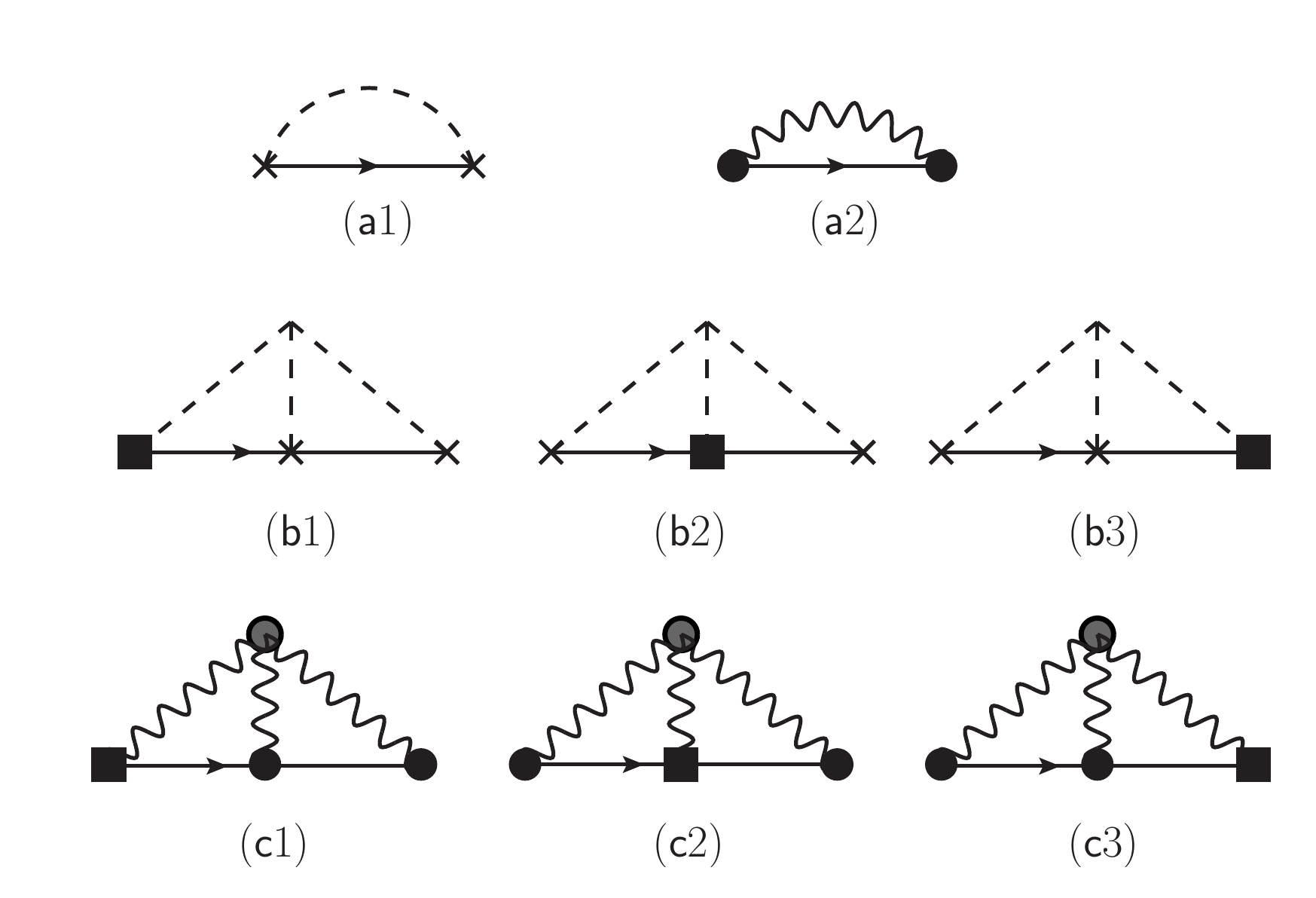}
\caption{(a) Self-energy in the standard self-consistent Born approximation
for electron-impurity (a1) and 
electron-phonon (a2) scattering. 
(b) Diagrams describing skew-scattering from impurities, and (c) diagrams
describing skew-scattering from phonons.
Dashed and wiggly lines indicate the impurity average and phonon propagator, respectively.  
The square box is the spin-orbit insertion due to both impurity and phonon potential.}
\label{fig1}
\end{center}
\end{figure}

The $T=0$ processes, as well as the dynamical side-jump and Elliott-Yafet, have been discussed in 
Ref.~[\onlinecite{toelle2014}]. In particular, skew scattering from impurities 
is described by the self-energy diagrams of Fig.~\ref{fig1}(b), 
together with the self-energy (a1) yielding the self-consistent Born approximation 
for the elastic scattering time.  In order to study finite (high) temperatures,
the self-energy (a2) as well as the skew-scattering from phonons via the self-energy diagrams
of Fig.~\ref{fig1}(c) have to be taken into account.

Just as for the second order self-energies [Fig.~\ref{fig1}(a)], there is a direct correspondence between 
the diagrams due to impurities [Fig.~\ref{fig1}(b)] and those due to phonon scattering [Fig.~\ref{fig1}(c)]. 
Such a correspondence appears in the high-temperature
limit, where the phonon dynamics become irrelevant, 
roughly speaking $\hat\varphi(\br,t) \rightarrow \varphi(\br)$ 
\footnote{The time-dependence of $\hat\varphi(\br,t)$ is due to the interaction picture.}.
To illustrate this further, consider the diagrams of Fig.~\ref{fig1}(b), 
with the impurity potentials---before averaging---replaced by the classical phonon 
field $g \varphi (\br)$.  The average is then performed using the equipartition law
\be
\label{gaussian_phonon}
\langle \varphi (\br_1)\varphi (\br_2)\rangle_T = k_B T \, \delta (\br_1 -\br_2) \, ,
\ee
where $\langle\dots\rangle_T$ denotes the classical average, and
\be
\label{three_phon_ave}
\langle \varphi (\br_1)\varphi (\br_2)\varphi (\br_3)\rangle_T =
- \Lambda (k_B T)^2  \, \delta  (\br_1-\br_2)\delta (\br_1-\br_3)
\ee
which follows from expanding the Boltzmann factor to first order with respect 
to the anharmonic term.  In the case of impurity scattering, the equivalent of 
the r.h.s.\ of Eq.~\eqref{gaussian_phonon}, assuming ``white-noise'' disorder, 
is given by $n_i v_0^2 \delta (\br_1 -\br_2)$, while the three-field
average results in $n_iv_0^3 \delta  (\br_1-\br_2)\delta (\br_1-\br_3)$. 
This suggests that one can obtain the high-$T$ results through the following correspondence:
\ber
n_iv_0^2 & \rightarrow & g^2k_B T  =\frac{1}{2\pi  N_0 \tau}\label{correspondence_1}\\
n_i v_0^3 & \rightarrow & -\Lambda g^3 (k_B T)^2=\frac{1}{2\pi N_0\tau} (-k_B T g\Lambda) \, .
\label{correspondence_2}
\eer
Note that $\tau$ here denotes the $T$-dependent electron-phonon scattering time, 
in contrast to $\tau_0$ due to elastic scattering from impurities; $-1/2\tau$ corresponds to 
the imaginary part of the retarded self-energy as derived
from (a2), with the result given in \eqref{correspondence_1} \cite{zimanbook2}.
Using \eqref{correspondence_1}, \eqref{correspondence_2} in Eq.~\eqref{sH0}, 
the skew scattering conductivity reads
\be
\label{ssT}
\sigma_{\rm ss}^{\rm sH}= -\left(\frac{\lambda k_F}{4}\right)^2
\frac{en}{m} \frac{\hbar \Lambda}{g} \, ,
\ee
which is, in particular, $T$-independent. Thus the $T$-dependence of the spin Hall conductivity
must interpolate between the two limiting expressions at low [see Eq.~\eqref{sH0}] 
and high [see Eq.~\eqref{ssT}] temperature.

In order to compare the order of magnitude of the two limits,
we use the standard relations valid for an isotropic metal, 
$v_0 \sim 1/2N_0$ (screened Coulomb impurities), $g^2 \sim 1/2N_0$, 
$\rho v_s^2 \sim N_0\epsilon_F^2$.  
For the ratio between the high- and low-$T$ conductivities we thus obtain
\be
\label{ratio}
\frac{\hbar \Lambda }{g\tau} \sim \frac{\gamma}{\epsilon_F\tau_0} \sim 0.1 \, ,
\ee
where, to be explicit, we assumed $\epsilon_F\tau_0/\hbar \approx 20$.
Note, however, that there might be a sign change as a function of temperature, depending
on the nature of the impurities, i.e., the sign of $v_0$, as well as on the sign of $g$.

A quantum field theoretical (Keldysh) calculation confirms the above results
up to a numerical prefactor in Eq.~\eqref{correspondence_2}.
Besides providing a solid basis for what has been obtained through simple and intuitive arguments,
we stress that such a calculation is necessary in order to study the full temperature 
range $0<T<T_D$.  

We now outline the Keldyh calculation in the high-$T$ regime, where the self-energy 
diagrams of Fig.~\ref{fig1}(c) acquire a transparent form \footnote{The self-energy [Fig.~\ref{fig1}(c)] for arbitrary $T$'s 
is fairly complicated, and will not be discussed here.}. In fact it is sufficient to consider 
the first one (now $\hbar=k_B=1$):
\begin{widetext}
\be
({\rm c}1): \; \Sigma_{{\rm ss},13}^{T} =
- \frac{\lambda^2}{4} \Lambda g^3 \sum_{i,j,k} \epsilon_{ijk}\,\sigma_j\,
(-i\nabla_1^G)_i (\nabla_1^D)_k \int_{2,4} G_{12}\, D_{14}\, D_{24}\, D_{34}\, G_{23} \, .
\ee
\end{widetext}
Here the $G$'s are $SU(2)$-covariant electron propagators \cite{gorini2010,raimondi2012,toelle2014},
while the $D$'s are free phonon propagators, both defined on the Keldysh contour.  
The arguments, written as subscripts, include both space and time, e.g., $1=({\br}_1,t_1)$.  
The notation $\nabla_1^G$ indicates that the gradient
acts only on the following $G$-function, and similarly for $\nabla_1^D$.
After analytical continuation \cite{rammer1986,*rammerbook2},
the $t$-integrals run from $-\infty$ to $+\infty$, and the Keldysh structure is carried 
by the $R,A,K$ propagator components.  In the high-$T$ regime, $T\gtrsim T_D$, 
we use $D^{<}\approx D^{>} \approx \frac{1}{2}D^K$
\footnote{In general $D^{>(<)} = [ D^K + (-) ( D^R-D^A ) ]/2$.}, with the result
\begin{widetext}
\be
\label{sigma1}
({\rm c}1): \; \left[\Sigma_{{\rm ss},13}^{T}\right]^{<(>)}
= - \frac{\lambda^2}{4} \sum_{i,j,k} \epsilon_{ijk}\,\sigma_j\,
(-i\nabla_1^G)_i (\nabla_1^D)_k
\int_2
\left( G^R_{12}\, G^{<(>)}_{23} + G^{<(>)}_{12}\, G^A_{23} \right) \mathbb{D}_{123} \, ,
\ee
\end{widetext}
where
\ber
\label{superD}
\mathbb{D}_{123} & = & \frac{\Lambda g^3}{4} \int_4 \, [ D^R_{14}\, D^K_{24}\, D^K_{34}\\ \nonumber
 & + & D^K_{14}\, D^R_{24}\, D^K_{34} + D^K_{14}\, D^K_{24}\, D^R_{34} ] \, . 
\eer
Equation~\eqref{sigma1} has the standard form due to the coupling to an
external field, whose role is here played by $\mathbb{D}$.  
Exploiting the fact that the phonon frequencies ($\sim \omega_D$) are small compared to 
$\omega \sim T$, which physically means that electron-phonon scattering is elastic, we obtain
\be
\mathbb{D}_{123} \approx -3\Lambda g^3 (k_BT)^2 \, ,
\ee
having restored here $k_B$ for easy comparison with Eq.~\eqref{correspondence_2}.
The only difference with the latter is a factor of 3, missed by the simple introductory argument.
The correct $T=0\rightarrow T>T_D$ correspondence for skew scattering thus reads
\be
\label{correspondence_3}
n_iv_0^3 \rightarrow -3\Lambda g^3 (k_BT)^2.
\ee
This yields at once
\be
\label{ssT3}
\sigma_{\rm ss}^{\rm sH}= -3\left(\frac{\lambda k_F}{4}\right)^2
\frac{en}{m} \frac{\hbar \Lambda}{g} \, ,
\ee
which is the central result of our work. Apart from the already mentioned factor of 3,
it confirms the heuristically obtained Eq.~\eqref{ssT}, and shows that the skew scattering
conductivity at high temperatures does not scale as the mobility, being rather $T$-independent.

\begin{figure}
\begin{center}
\includegraphics[width=\columnwidth]{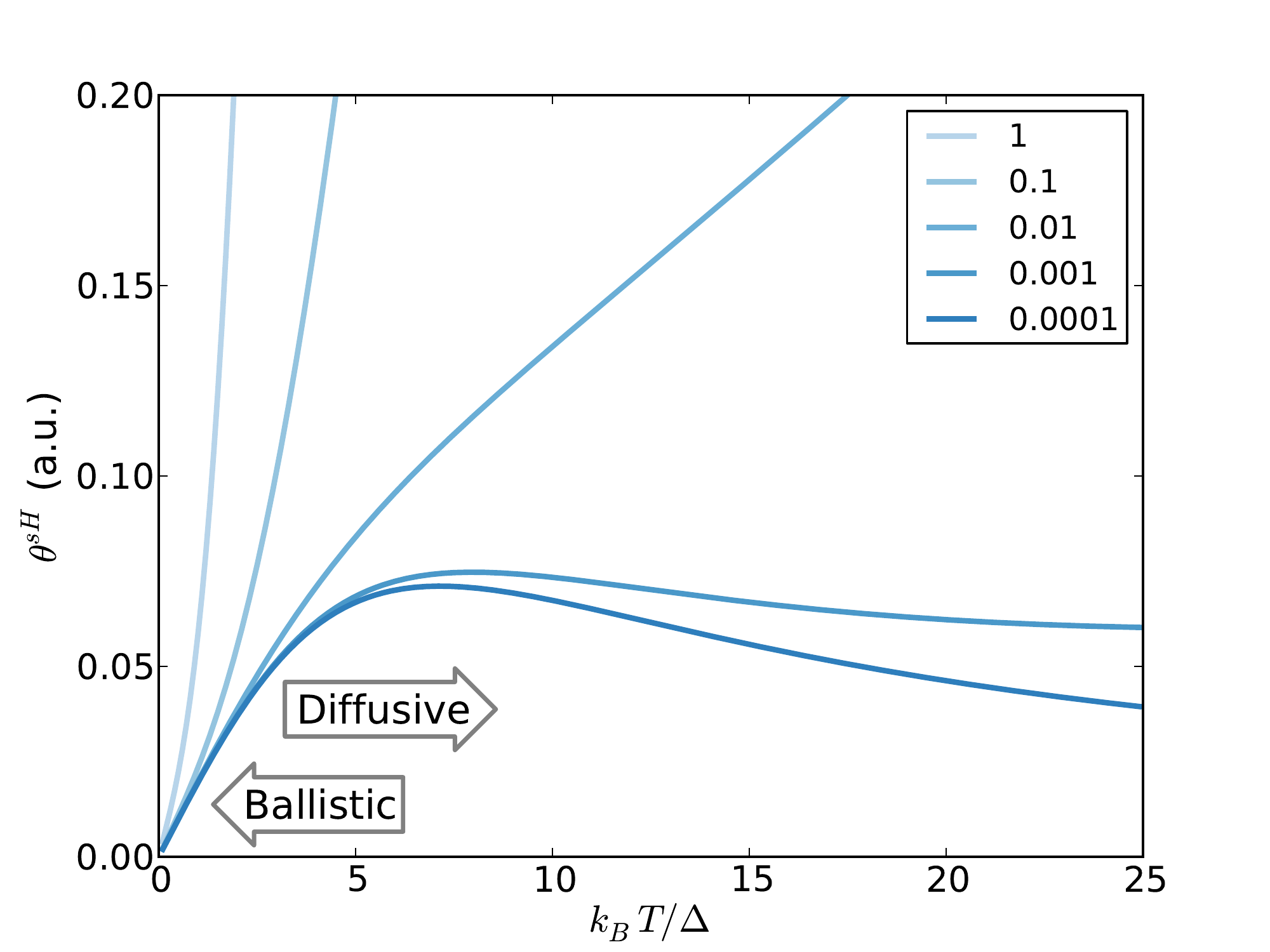}
\caption{Qualitative plot of the $T>T_D$ spin Hall angle $\theta^{\rm sH}$ 
as a function of $k_BT$, measured in units of the intrinsic spin-orbit splitting,
for the paradigmatic case of a Rashba-like system.
The spin Hall conductivity is given by Eq.~\eqref{sHfull}, and we set 
$\lambda/\lambda_F \approx 10^{-1}$ \cite{toelle2014}. Darker (lighter) curves are for 
weaker (stronger) extrinsic conductivities,
$(\sigma^{\rm sH}_{\rm sj}+\sigma^{\rm sH}_{\rm ss})/(e/8\pi\hbar) = 10^{-4} \dots 1$.
}
\label{fig2}
\end{center}
\end{figure}

We stress that the current interpretation of (inverse) spin Hall experiments is, however,
based on the ``scaling-as-mobility'' assumption \cite{hankiewicz2006,vila2007,vignale2010,niimi2011,isasa2014}.
Equation~\eqref{ssT3} shows that a more careful analysis seems to be required,
and has important consequences for the spin Hall angle 
$\theta^{\rm sH}\equiv e \sigma^{\rm sH}/\sigma_D$. 
 As shown in Fig.~\ref{fig2}, the spin motion becomes diffusive for $\Delta < \hbar/\tau \sim k_BT \sim10^{-2}$ eV
and ballistic for $\Delta > k_BT$, with $\Delta=2\alpha k_F$ the intrinsic splitting.
If a nonlinear or decreasing behavior of $\theta^{\rm sH}$ is observed, we deduce that $k_BT>\Delta$
and $(\sigma^{\rm sH}_{\rm sj} + \sigma^{\rm sH}_{\rm ss})/(e/8\pi\hbar)\ll1$ (extrinsic effects
are much weaker than the intrinsic ones).  If, on the other hand, a linear behavior
is observed, no conclusion can be reached by simply looking at the $T$-dependence,
since there are two possibilities:
(i)  $k_BT > \Delta$ and the extrinsic and intrinsic effects are comparable (light blue
curves in Fig. 2);
(ii) $k_BT < \Delta$ and nothing can be said about the relative strength of extrinsic
and intrinsic mechanisms (all curves, i.e., for different parameter values, look the same).

The relative importance of phonon vs.\ impurity skew scattering is obtained
by comparing the self-energies (b1) and (c1), yielding
\be
\label{estimate}
{\Sigma_{\rm ss}^T}/{\Sigma_{\rm ss}^0} \sim -\gamma ({\tau_0}/{\tau}) ({k_BT}/{\epsilon_F}) \, .
\ee
In a metal at room temperature we have $k_B T/\epsilon_F\sim 10^{-2}$, setting as threshold 
for the dominance of phonon skew scattering $\tau_0\gtrsim 10^2\tau$.

In general, the $T=0 \rightarrow T>T_D$ correspondence lets us immediately turn known
$T=0$ results into their $T>T_D$ counterparts.  For example,
the full expression for the high-$T$ spin Hall conductivity and current-induced spin polarization
\cite{ivchenko1978,*vasko1979,*levitov1985,*aronov1989,*edelstein1990,*ganichevreview2011}
due to intrinsic Bychkov-Rashba coupling and extrinsic {\it dynamical} spin-orbit interaction
is structurally identical to the $T=0$ expressions appearing in Ref.~\cite{raimondi2012}.
Explicitly for a 2D homogeneous bulk system
\be
\label{sHfull}
\sigma^{\rm sH} = \frac{1}{1+\tau_s/\tau_{\rm DP}}\left(\sigma^{\rm sH}_{\rm int} 
+ \sigma^{\rm sH}_{\rm sj} + \sigma^{\rm sH}_{\rm ss}\right)
\ee
where $\sigma^{\rm sH}_{\rm int}=(e/8\pi\hbar)(2\tau/\tau_{\rm DP})$ is the intrinsic part of the
spin Hall conductivity, and
\be
\frac{1}{\tau_{s}} = \frac{1}{\tau}\left(\frac{\lambda k_F}{2}\right)^4,\;
\frac{1}{\tau_{\rm DP}} = \frac{1}{2\tau}\frac{(\Delta\tau/\hbar)^2}{
[(\Delta\tau/\hbar)^2+1]}
\ee
are, respectively, the Elliott-Yafet and Dyakonov-Perel spin relaxation rates.  
Furthermore, the current-induced spin polarization
 ``conductivity'', ${\cal P}$, is given by
\be
{\cal P} = -\frac{2m\alpha}{\hbar^2}\frac{1}{1/\tau_s+1/\tau_{\rm DP}}
\left(\sigma^{\rm sH}_{\rm int} 
+ \sigma^{\rm sH}_{\rm sj} + \sigma^{\rm sH}_{\rm ss}\right) \, ,
\ee
This phenomenon, together with its inverse \cite{ganichev2002, rojassanchez2013, shen2014},
is intimately related to the spin Hall effect \cite{gorini2008, shen2014}
and can be similarly exploited for spin-to-charge conversion 
\cite{ganichev2002, rojassanchez2013, shen2014}.

We conclude by discussing future perspectives and certain limitations of our approach.
First, the anharmonic term \eqref{phonon_anharmonic} was handled via an ``$s$''-wave approximation,
ignoring the tensor structure of $\Lambda$ as well as any details of the generally anisotropic
phonon-phonon coupling: these, however, are not expected to
{\it qualitatively} modify our conclusions concerning the $T$-dependence.
The same is true when other phonon modes are included, provided
their typical frequencies are smaller than $k_B T/\hbar$.

Second, $\phi^4$ (and higher) anharmonicities, formally necessary to stabilize the system, could
also be considered. These have their $T=0$ parallel in the T-matrix resummation of skew scattering.  
However, whereas the latter does not add qualitative new features to the physics 
described by the diagrams of Fig.~\ref{fig1}(b), higher anharmonicities could.  
Roughly speaking, any additional phonon line connected to the anharmonic 
vertex in the diagrams of Fig.~\ref{fig1}(c) should contribute a further $k_BT$ factor in the $T>T_D$ regime,
as well as modifying the prefactor of ``3'' missed by the simple introductory arguments.
This would further increase the importance of phonon skew scattering at high $T$'s,
possibly implying a $T$-behavior of $\sigma^{\rm sH}_{\rm ss}$ opposite to that 
of the mobility.  Indeed, it would be highly desirable to develop a more detailed theory of phonon 
scattering, in analogy with the $T=0$ treatment by Fert and Levy \cite{fert2011}, 
as well as to elucidate the role of Umklapp processes.

Third, band nonparabolicities could be relevant since they modify, in particular, 
the side-jump mechanism, and thus possibly its $T$-dependence.  
Finally, and probably most importantly, 
the intermediate temperature regime, $0<T<T_D$, needs to be properly investigated.
We stress that our Keldysh approach gives an expression for the self-energy [Fig.~\ref{fig1}(c)] formally 
valid for all temperatures. However, at lower $T$'s the interplay between interactions, 
impurity scattering and phonons can have important consequences \cite{schmid1973}. 
We expect that our results will stimulate further
(much needed) work in these directions of highest experimental relevance.

C.~G. and U.~E. acknowledge financial support from the Deutsche Forschungsgemeinschaft through SFB 689 and TRR 80, respectively.

\bibliography{paper_pss_biblio}

\end{document}